\documentstyle[twocolumn,epsfig,eqsecnum,aps]{revtex}

\begin{document}
\draft
\preprint{PAL 00-1}
\title{Energy Spectra, Altitude Profiles and Charge Ratios of Atmospheric 
Muons}
\author{S. Coutu, J. J. Beatty and M. A. DuVernois}
\address{
Departments of Physics and of Astronomy and Astrophysics, 104 Davey 
Laboratory, The Pennsylvania State University, University Park, PA 16802
}
\author{S. W. Barwick and E. Schneider}
\address{Department of Physics, University of California at Irvine, Irvine, 
CA 92717}

\author{A. Bhattacharyya, C. R. Bower and J. A. Musser}
\address{Department of Physics, Swain Hall West, Indiana University,
Bloomington, IN 47405}

\author{A. Labrador, D. M\"uller, S. P. Swordy and E. Torbet\footnote{Present 
address: Department of Physics, Broida Hall, University of California at
Santa Barbara, Santa Barbara, CA 93106}}
\address{Enrico Fermi Institute and Department of Physics, 933 E. 56$^{th}$
St., University of Chicago, Chicago, IL 60637}

\author{C. Chaput, S. McKee, G. Tarl\'e and A. D. Tomasch}
\address{Department of Physics, Randall Laboratory, University of Michigan,
500 E. University, Ann Arbor, MI 48109-1120}

\author{S. L. Nutter}
\address{Department of Physical Sciences, Eastern New Mexico University,
Portales, NM 88130, USA}

\author{G. A. deNolfo\footnote{Present address: Department of Physics, Downs 
Laboratory, California Institute of Technology, Pasadena, CA 91125}}
\address{Department of Physics, Washington University, St. Louis, MO 63130}

\date{\today}
\maketitle
\begin{abstract}
We present a new measurement of air shower muons made during atmospheric 
ascent of the High Energy Antimatter Telescope balloon experiment. The muon 
charge ratio $\mu^+ / \mu^-$ is presented as a function of atmospheric depth 
in the momentum interval 0.3--0.9~GeV/c. The differential $\mu^-$ momentum
spectra are presented between 0.3 and $\sim$50 GeV/c at atmospheric depths
between 13 and 960~g/cm$^2$. We compare our measurements with other recent
data and with Monte Carlo calculations of the same type as those used in 
predicting atmospheric neutrino fluxes. We find that our measured $\mu^-$ 
fluxes are smaller than the predictions by as much as 70\% at shallow
atmospheric depths, by $\sim$20\% at the depth of shower maximum, and are
in good agreement with the predictions at greater depths. We explore 
the consequences of this on the question of atmospheric neutrino production.
\end{abstract}
\pacs{PACS numbers: 96.40.Tv, 14.60.Pq, 14.60.Ef}

\narrowtext

\section{Introduction}
\label{intro.sec}

Measurements of the flux of atmospheric neutrinos by large underground 
detectors have consistently disagreed with theoretical predictions, a fact 
that has been interpreted in terms of possible neutrino oscillations.
The most compelling evidence thus far comes from the 
SuperKamiokande experiment \cite{fuk:osc}. The discrepancies between 
experiment and theory are well beyond the statistical uncertainties in the 
measurements. The correct interpretation of the effect requires a detailed
understanding of neutrino production in the atmosphere. 

Model predictions \cite{bar:nu,lee:nu,hon:nu,kaw:nu,agr:nu} of the 
{\it absolute} flux of neutrinos produced in air showers are uncertain 
due to a number of difficulties: one problem is the normalization of the 
primary cosmic-ray flux, for which measurements vary by $\pm$15\% or more. 
Because of this, different assumptions are made by different authors. For 
instance, the model of Honda {\it et al.} \cite{hon:nu}, used as the starting 
point in the analyses of the SuperKamiokande collaboration, assumes a primary 
cosmic ray flux normalization that is in excess of the current measurements 
by balloon experiments \cite{bes:pfl}, by as much as $\sim$20\% in the energy 
range beyond about 5~GeV, which is relevant to the production of sub- to 
multi-GeV atmospheric neutrinos. In contrast, the model of the Bartol group 
\cite{agr:nu,gai:nu} uses a primary spectrum in good agreement with the most 
recent measurements. Different assumptions are also made about the details of 
particle production in atmospheric interactions, and about the treatment of 
geomagnetic effects (see Ref.~\cite{gai:rat}). Thus, while the authors of these
calculations expect that the predictions of absolute neutrino fluxes might 
have a total uncertainty of about 18\% \cite{agr:nu}, it seems fortuitous that
the predicted neutrino rates have been found to agree with each other within 
about 10\%. It should also be noted that in the neutrino oscillation analysis 
of the SuperKamiokande collaboration, an additional normalization factor of 
1.16 in the cosmic ray intensity is introduced to obtain the best oscillation 
fit to the measured muon and electron neutrino rates as a function of zenith 
angle. 

Neutrino production in the atmosphere is closely coupled to muon
production, as both types of particles are produced together in pion and kaon 
decays, and as some of the muons decay to produce further neutrinos. 
Monte Carlo simulations of neutrino production naturally predict the 
spectra of muons as a function of atmospheric depth. Therefore, a detailed
experimental test of the predictions is possible through measurements of the
intensity of muons at different depths in the atmosphere, for instance
with a balloon-borne particle detector during atmospheric ascent.
If such a detector includes a magnet spectrometer, separate measurements of
the $\mu^-$ and $\mu^+$ fluxes are possible. This is relatively easy
for negative muons, as negatively charged particles other than electrons are
rare in air showers (and electrons are easily rejected), while non-interacting 
protons generate a large background for positive muons above $\sim$1~GeV. 
Such measurements have been made by a number of instruments, including the 
MASS \cite{mas:mu,mas2:mu}, HEAT \cite{sch:mu,heat:mu}, 
CAPRICE \cite{boe:mu}, and IMAX \cite{kri:mu} magnet spectrometers and 
older, less sensitive balloon payloads.

Here we describe measurements made with the High Energy Antimatter Telescope 
(HEAT), a balloon instrument optimized for the study of high-energy cosmic-ray 
electrons and positrons. HEAT uses several complementary particle 
identification techniques, which are also well suited to 
the identification of muons during atmospheric ascent of the balloon. 
This instrument was flown twice, in 1994 and 1995. Preliminary results from
the first flight on relative abundances of $\mu^+$ and $\mu^-$ were reported 
previously \cite{sch:mu}. The present paper describes measurements made during
the second balloon flight: the relative abundances of $\mu^+$ and $\mu^-$ at 
momenta between 0.3 and 0.9~GeV/c, and the differential spectra of $\mu^-$ at 
momenta between 0.3 and 50~GeV/c, at atmospheric depths between 13 and 
960~g/cm$^2$. We compare the results with other measurements and with 
calculations made with a modified version of the TARGET Monte Carlo 
algorithm \cite{agr:nu}, developed by the Bartol group to predict
atmospheric neutrino rates. 

\section{Muon Identification and Backgrounds}
\label{muid.sec}

The HEAT instrument is described in detail elsewhere \cite{bar:nim}.
It combines a superconducting magnet spectrometer (using
a drift-tube tracking hodoscope), a time-of-flight (TOF) system, a
transition-radiation detector (TRD) and an electromagnetic calorimeter (EMC).
For its second flight, it was launched from Lynn Lake, 
Manitoba, Canada, at a vertical geomagnetic cutoff rigidity of well below 
1~GV. The flight took place on 23 August 1995, approximately at minimum
solar activity. The ascent from an atmospheric overburden at the
ground of 960~g/cm$^2$ to a float altitude at 3~g/cm$^2$ required 3.1~hours, 
during which charged atmospheric secondary particles were detected and 
recorded. Near the end of ascent, at about 13~g/cm$^2$, the instrument trigger
configuration was changed to commence preferential measurements of 
electrons and positrons, affecting the ability to measure absolute muon
intensities at shallower depths. Therefore we report measurements of the muon
charge ratio at all atmospheric depths, but measurements of absolute muon
intensities only at depths greater than 13 g/cm$^2$.

In identifying muon events, TRD information is not used. The TOF
system measures the velocity of the particle $\beta = v/c$ with a resolution
of $\sigma_\beta = 0.15$, permitting complete rejection of upward-going
particles; in addition, the amount of light generated in the scintillation
counters of the TOF system is used to measure the magnitude of the particle's 
electric charge $Ze$ with a resolution of $\sigma_Z = 0.11$, permitting the 
unambiguous identification of singly-charged particles. 
The magnet spectrometer measures the sign of the particle's charge from the 
direction of the deflection in a magnetic field of about 1~T, with a field
integral over the particle's trajectory of $\int B \cdot dl \sim 4.2$~kGm.
The magnetic rigidity $R = pc/Ze$ of the particle is determined from the 
amount of deflection; at rigidities between 0.3 and 0.9~GV, the resolution 
achieved is $\sigma_R = 0.08$ to 0.11~GV.
The EMC records the 
pattern of energy deposits in 10 scintillation counters, each topped by a 0.9
radiation length-thick lead sheet. In each layer, the energy deposited
is measured in units of the energy loss by a vertical minimum-ionizing
particle (MIP). A shower sum is obtained by adding the signals from the 10
scintillators.

\begin{figure}
\epsfig{file=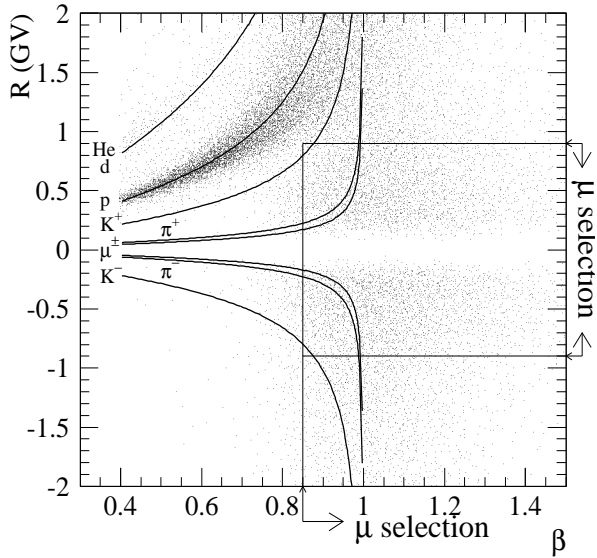,width=8.5cm,height=8.0cm}
\caption{Distribution of particle rigidity $R$ as a function of velocity
$\beta = v/c$ for ascent data.} \label{rvsb}
\end{figure}

\begin{figure}
\epsfig{file=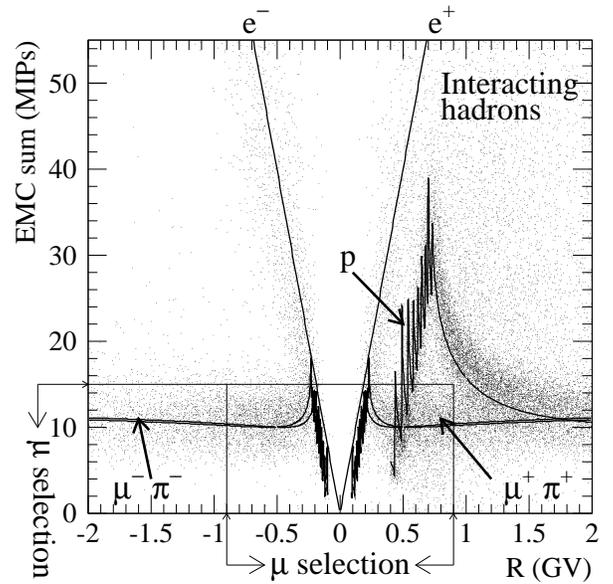,width=8.5cm,height=8.0cm}
\caption{Distribution of shower sum in the EMC as a function of particle
rigidity $R$ for ascent data.} \label{svsr}
\end{figure}

The selection of muon events proceeds in three steps.
First, a high-quality spectrometer track and measurement of
the rigidity $R$ are required, as described in Ref.~\cite{heat:mu}. 
Also required are down-going and singly-charged TOF characteristics.
Second, as shown in Figure~\ref{rvsb}, the 
distribution of reconstructed particle rigidity $R$ is studied as a function
of velocity $\beta$, and compared with ideal curves
$R = (m/Z) \beta / \sqrt{(1-\beta^2)}$ where $m$ is the mass and $Z$ the 
charge of the particle. Clearly identifiable in the figure are populations due 
to He (and d), p, $\mu^+$ and $\mu^-$ events. $\pi^\pm$ and some $K^\pm$ events
cannot be distinguished from the $\mu^\pm$ populations, and are a small 
background to the muon signal. We select $\mu$-like events by requiring: 1) 
$\beta \geq 0.85$ and 2) $0.3 \leq \left| R \right| \leq 0.9$~GV, for the 
low-energy $\mu^+/\mu^-$ ratio, and 2) $R \leq -0.3$~GV for the $\mu^-$
energy spectra. For $R\geq 0.9$~GV, $\mu^+$ events become
indistinguishable from non-showering hadrons (mostly protons), so that
$\mu^+$ spectra are not measured with this instrument.
Third, we study the behavior of the shower sum measured by the EMC 
as a function of particle rigidity $R$, as shown in Figure~\ref{svsr}.
Ideal curves for electrons and positrons are obtained by assuming a simple 
linear relationship between EMC sum (proportional to energy) and $R$. For
heavier particles, a calculation of EMC sum is made by integrating
Bethe-Bloch energy losses within the EMC scintillators. Populations
of events due to e$^\pm$, p and $\mu^\pm/\pi^\pm/K^\pm$ are identifiable.
By augmenting the selection criteria of Figure~\ref{rvsb} with the requirement 
that particles not shower in the EMC (EMC sum $\leq$ 15), e$^\pm$ events 
with $\left| R \right| > 0.3$~GV are rejected. The low-rigidity proton events
that range out in the EMC and would appear to contaminate the $\mu^+$
population are rejected by the $\beta \geq 0.85$ requirement of 
Figure~\ref{rvsb}.

With these criteria, we achieve essentially complete rejection of 
electron events, but
there remains a small background to the muon signal due to pions and kaons.
We estimate from Monte Carlo simulations of air showers based on CERN's 
GEANT-FLUKA algorithms \cite{fas:flu} that the $\pi^\pm$
flux at a depth of 13~g/cm$^2$ is only 2\% that of $\mu^\pm$, in agreement
with another calculation \cite{ste:sec} -- with $K^\pm$ fluxes at a much lower
level -- and that this further decreases with increasing atmospheric depth.
Moreover, only pions that do not interact can be mistaken for muons,
which occurs 39\% of the time, so that the background to the muon measurement
due to atmospheric pions is only about 0.7-0.9\% near float, decreasing to
less than 0.4\% at depths greater than 300~g/cm$^2$. Such a small background
is not corrected for here. Occasionally, cosmic-ray interactions in the 
instrument result in $\pi^\pm$ production, at an even more modest level than
the atmospheric pion background. GEANT-based simulations indicate that only
0.04\% of proton-induced events yield a misidentification as a $\mu^-$.
As the payload slowly rotated throughout the flight (as determined with a 
solar sensor attached outside the gondola), and did not align itself with the
Earth's magnetic field, possible geomagnetic East-West asymmetries are
averaged out. Furthermore, such asymmetries in the primary proton flux are
only expected at momenta near the geomagnetic cutoff, which for the Lynn Lake
flight is well below the energies of interest here.

\section{Results and Discussion}
\label{disc.sec}

\subsection{Muon Charge Ratio}

\begin{figure}
\epsfig{file=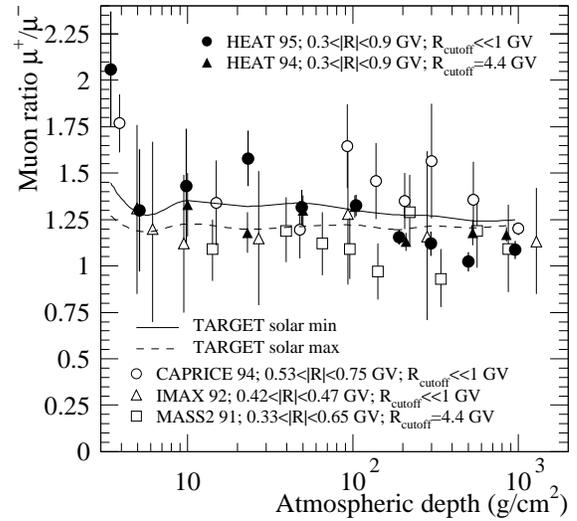,width=8.0cm}
\caption{Measured muon ratio $\mu^+/\mu^-$ as a function of atmospheric 
depth. The curves are calculations with the TARGET algorithm. The
HEAT data are for $0.3\leq\left|R\right|\leq0.9$~GV. \label{ratio}}
\end{figure}

The number of low-energy $\mu^+$ and $\mu^-$ events detected as a function of 
atmospheric
depth and the $\mu^+/\mu^-$ ratio are summarized in Table~\ref{mupmum}. The
$\mu^+/\mu^-$ ratio is shown as a function of atmospheric depth in
Figure~\ref{ratio} (labeled HEAT 95), together with the measured ratio from 
the first HEAT flight~\cite{sch:mu} (labeled HEAT 94) and other recent 
measurements~\cite{boe:mu,kri:mu,bas:mu}. 
As noted on the figure, the various analyses have used different magnetic 
rigidity ranges, and moreover, the measurements were made at different solar 
epochs and different geomagnetic rigidity cutoffs R$_{\rm cutoff}$, so that 
the data are not truly directly comparable.
The two HEAT measurements of the muon ratio are the most statistically
significant, and are essentially consistent with each other and with other 
measurements, within the appreciable errors. Based on the HEAT measurements,
no clear correlation of the muon ratio with geomagnetic cutoff rigidity
is observed. The depth dependence of the charge ratio seems to be essentially 
flat, although a slight decrease from about 1.3-1.4 at high altitudes 
(3-50~g/cm$^2$) down to about 1.1 at the ground
cannot be excluded. 
We note that the charge ratio measured by both HEAT 95 and CAPRICE 94 
at small depths appears anomalously high. This effect is not understood at
present.
Also shown on Figure~\ref{ratio} are
calculations with the TARGET algorithm~\cite{agr:nu}
(widely used for neutrino flux calculations), for conditions of solar minimum 
and maximum activity. Both HEAT flights occurred under essentially 
solar-minimum conditions. 
The TARGET calculations have been made for average primary fluxes at solar 
minimum and maximum, which may not exactly represent the actual spectrum at 
the time of the flight. The calculations are for a location with no 
geomagnetic cutoff, and are intended for comparison with the HEAT 95 data
only. The agreement between the HEAT measurements and the solar-minimum TARGET 
calculation appears to be fairly good.
Note that the fluctuations in the simulated distributions at shallow 
atmospheric depths are statistical, and indicative of the small number of 
muons having been produced in the air showers at the highest altitudes.

\subsection{Energy Spectra of Negative Muons}

The absolute intensity of $\mu^-$, in a rigidity interval $\Delta R$ with
an average rigidity $\overline{R}$, at atmospheric depth $d$ is obtained with:

$$ j_\mu (d,\overline{R}) = {N_\mu \over \Delta t \epsilon_t \epsilon_l
\epsilon_\theta \epsilon_{dt} \Delta R (\Omega A) 
\epsilon_{scan} \epsilon_{acc}}, $$

\noindent
where $N_\mu$ is the number of $\mu^-$ events recorded at $(d,\overline{R})$,
$\Delta t$
is the time spent at depth $d$, $\epsilon_t$ is the live time fraction,
$\epsilon_l$ is an event-transmission loss correction, $\epsilon_\theta$ is
a correction to account for the fact that the muon flux is increasingly less
isotropic deeper in the atmosphere, $\epsilon_{dt}$ is the efficiency of the
basic event cleanliness criteria applied to the drift-tube hodoscope track,
$(\Omega A)$ is the geometrical factor, $\epsilon_{scan}$ is a ``scanning
efficiency'' correction (described below), and $\epsilon_{acc}$ is the 
muon acceptance efficiency. The acceptance of the instrument decreases rapidly
for particles incident at a zenith angle greater than about 25$^\circ$, so that
particle intensities reported here are essentially for vertical incidence.

Both $\overline{R}$ and $\Delta R$ are weighted to account for the 
details of the energy spectrum, according to: 

$$\overline{R} = {\int_{R_i}^{R_j} R f(R) dR \over \int_{R_i}^{R_j} f(R) dR}
\;\;\;\;\;\;
\Delta R = {\int_{R_i}^{R_j} f(R) dR \over f(\overline{R}) } $$

\noindent where $f(R) \propto R^{-\alpha}$ is the rigidity power-law spectrum, 
with spectral index $\alpha$ varying between $-0.56$ and 3.5 depending on both 
$R$ and $d$ ($\alpha$ is 
experimentally determined from the spectra before any of the normalization
corrections are applied). $\Delta t$ is measured with an on-board clock,
$\epsilon_t$ is determined using on-board scalers which count clock cycles 
while the instrument is available for a trigger or busy processing an event,
and $\epsilon_l$ is determined by careful accounting of event numbers 
generated on board compared to events successfully transmitted. 
$\epsilon_\theta$ is calculated using a standard prescription
\cite{zen:cor}, where the zenith dependence of the muon flux is taken
to be $\cos ^{n(d)} \theta$, with the exponent $n$ a function of atmospheric
depth $d$; we have used $n(d)=(d/1030 {\rm g/cm^2}) \times n_{sea level}$
with $n_{sea level}=2$.
$\epsilon_{dt}$ is obtained by a careful accounting of the
number of events recorded compared to the number of events with a successful
minimal track reconstruction. $(\Omega A)$ and $\epsilon_{acc}$ are determined
with the aid of a GEANT-based simulation of the response of the HEAT
instrument. $\epsilon_{scan}$ is a correction factor introduced based
on the visual scanning of several hundred events to account for residual
differences between the reconstruction efficiency of real events compared to
that of simulated ones, and is found to be $\epsilon_{scan} = (0.9\pm 0.1)$.
The various parameters described above are given in Tables~\ref{reffic} and
\ref{deffic}.

\begin{figure}
\epsfig{file=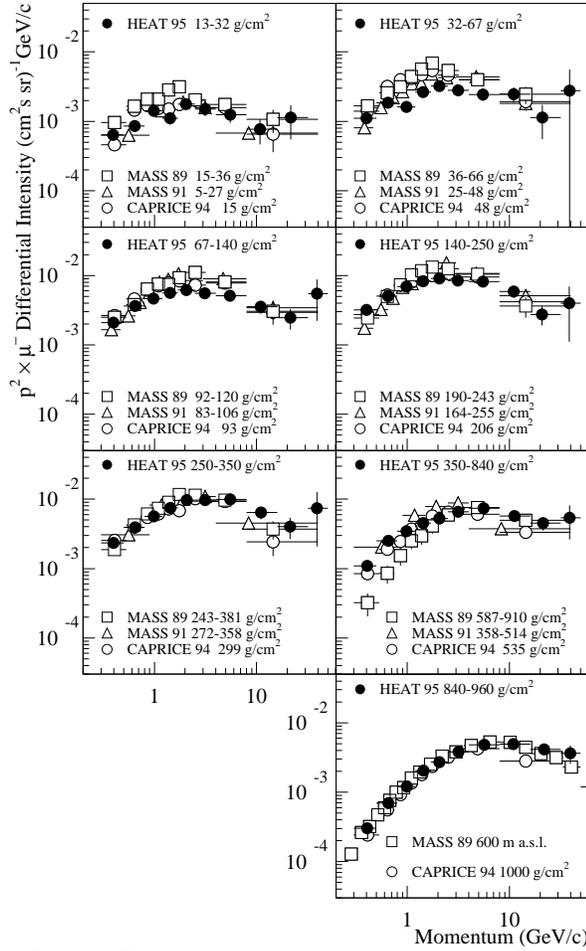,width=8cm}
\caption{Differential $\mu^-$ momentum spectra as a function of atmospheric 
depth. The HEAT measurements are compared with the MASS
and CAPRICE measurements for qualitative purposes only.
\label{ffluxes}}
\end{figure}

The final $\mu^-$ intensities as a function of momentum for various 
atmospheric depths
are shown in Figure~\ref{ffluxes} and given in Table~\ref{fluxes}. 
Also shown in the figure are the measurements of the MASS \cite{mas:mu,mas2:mu}
and CAPRICE \cite{boe:mu} experiments. The HEAT sample of 10327 $\mu^-$ events
collected during ascent is to be compared with the MASS samples of 2893 events
(1989 flight) and 4471 events (1991 flight) 
and the CAPRICE sample of 4627 events. Although the 1989 MASS measurements 
were also made in Northern Canada (from Prince Albert, Saskatchewan), the 
flight occurred at a different solar epoch (1989), at the time of a 
significant Forbush decrease. The 1991 MASS measurements were made from
Fort Sumner, New Mexico. The CAPRICE data were collected at Lynn Lake,
in 1994, and so are more directly comparable with our measurements. 
The general level of agreement between the data sets should be noted. 

\subsection{Comparison with Model Calculations}

\subsubsection{One-Dimensional TARGET Algorithm}

\begin{figure}
\epsfig{file=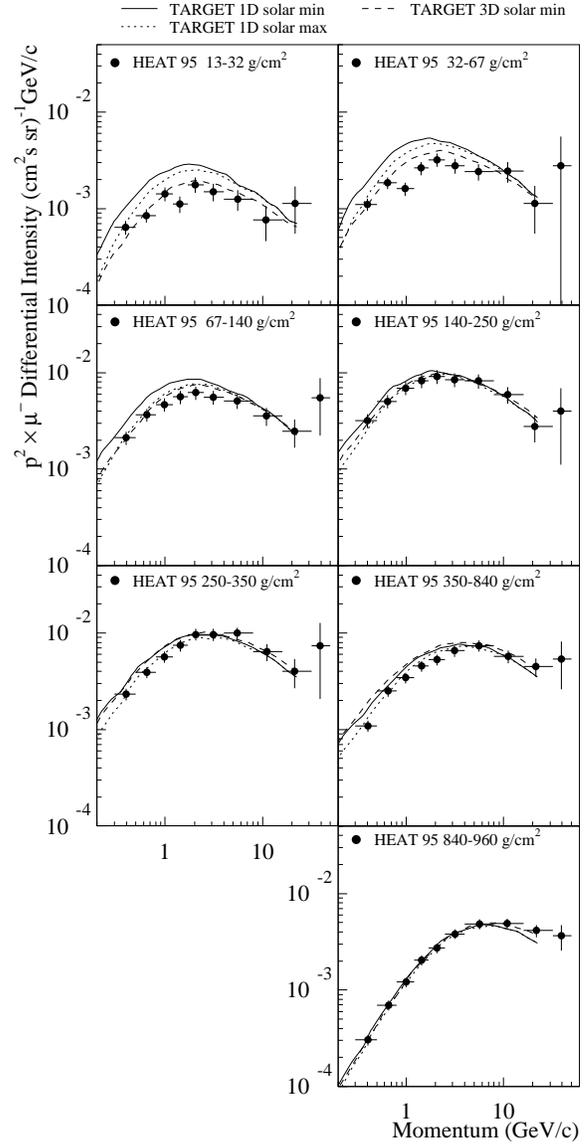,width=8cm}
\caption{Differential $\mu^-$ momentum spectra measured by HEAT
for various atmospheric depths. The curves are predictions 
of the 1D TARGET algorithm at solar minimum (solid curves) or solar
maximum (dotted curves) conditions, and the 3D TARGET algorithm at
solar minimum (dashed curves), respectively.
\label{flxtarget}}
\end{figure}

\begin{figure}
\epsfig{file=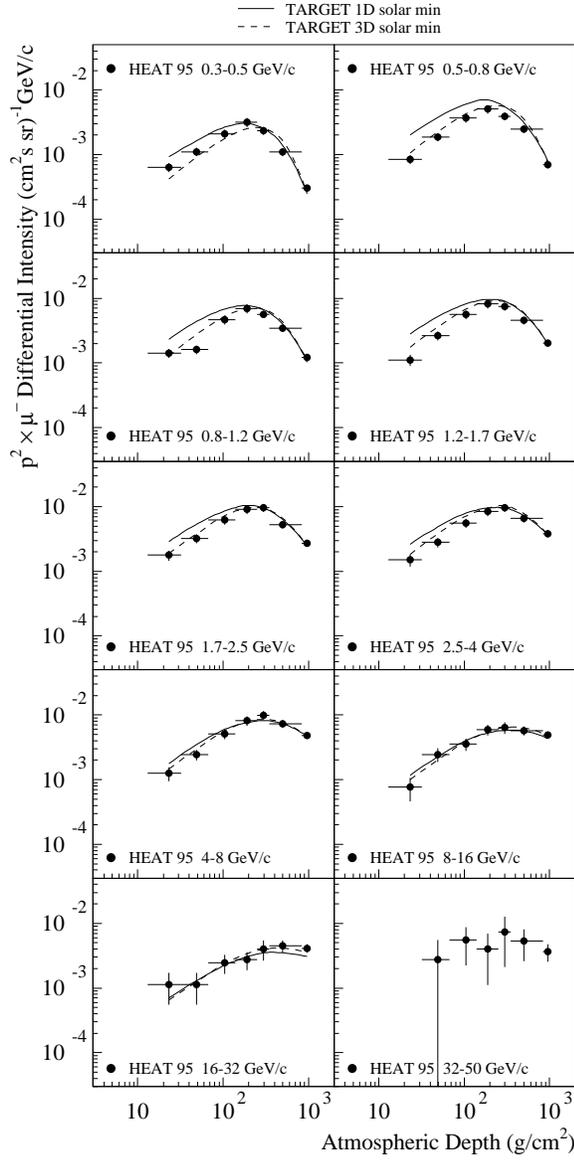,width=8cm}
\caption{$\mu^-$ growth curves measured by HEAT for different momentum
intervals. The curves are predictions of the 1D TARGET algorithm (solid
curves) and 3D TARGET algorithm (dashed curves), respectively, for solar 
minimum conditions.
\label{grotarget}}
\end{figure}

In Figure~\ref{flxtarget}, we compare the HEAT measurements reported here
with predictions of the TARGET algorithm \cite{agr:nu}, for conditions
of solar minimum and maximum activity, shown as solid and dotted curves,
respectively. (The solar-minimum curves are the ones of interest here, but 
the solar maximum curves are shown as well to illustrate the extent of the 
effect of the solar cycle on muon production.) These curves are obtained
with the standard TARGET algorithm, which simulates vertically incident cosmic 
rays, and which follows the development of the air shower in one dimension 
only. No corrections for geomagnetic effects are made. This is the 
algorithm developed by the Bartol group and used in predicting underground 
neutrino rates from atmospheric sources \cite{agr:nu,gai:nu}. In 
Figure~\ref{grotarget}, we show the $\mu^-$ growth curves for different 
momentum intervals, also compared with the 1D TARGET-based predictions for
solar minimum conditions (solid curves). The calculations were not made for the
highest momentum bin. 

There is general similarity between
the experimental and simulated distributions, with some notable differences. 
For instance, the predictions are significantly in excess of the measurements 
below 4~GeV/c
at atmospheric depths between 13 and 250~g/cm$^2$. The ratio of simulated
to measured intensity varies from $1.2 \pm 0.2$ near shower maximum at
200~g/cm$^2$ to $1.7 \pm 0.3$ at depths between 13 and 140~g/cm$^2$. 
At depths beyond shower maximum, or at momenta greater than
4~GeV/c, the simulations agree very well with our measurements.
A similar trend was found by the CAPRICE collaboration \cite{boe:mu}:
simulations predict more $\mu^-$ events than they measure below about 
1~GeV/c, but they find that the ratio of simulated to measured intensity is 
greatest at shower maximum, with a value of $1.8 \pm 0.1$. 

\subsubsection{Three-Dimensional TARGET Algorithm}

In an attempt to understand the origin of the discrepancy between the
predicted and measured muon intensities, a new version of the TARGET
algorithm was produced in collaboration with T. Gaisser and T. Stanev of
the Bartol Research Institute. In this, three dimensional air shower
development effects are taken into account, and the primary cosmic ray
arrival direction is sampled isotropically, rather than assuming vertical
incidence. Geomagnetic effects are not yet included in the calculations. 
Figure~\ref{flxtarget} also shows the $\mu^-$ momentum spectra at various
atmospheric depths obtained with the 3-dimensional TARGET algorithm 
(dashed curves), for solar minimum conditions. The 3D calculations are in 
substantially better agreement with the data than the 1D calculations. 
In Figure~\ref{grotarget}, the measured $\mu^-$ growth curves
are also compared with 3D TARGET predictions (dashed curves). Here again the 
3D predictions are a more adequate representation of the data.

\begin{figure}
\epsfig{file=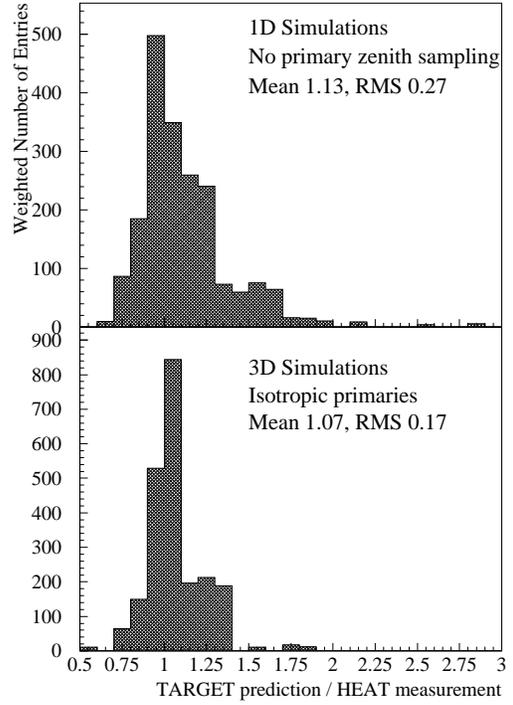,width=7cm}
\caption{
Distribution of the ratios of predicted to measured $\mu^-$ intensities,
for 1D TARGET calculations (top panel) and 3D TARGET calculations (bottom
panel), respectively. 
Each ratio is weighted by the square of the error on the ratio derived
from the experimental error on the intensity.
\label{residuals}}
\end{figure}

Figure~\ref{residuals} shows distributions of the ratios of predicted to 
measured $\mu^-$ intensities, for 1D TARGET calculations (top panel) and 
3D TARGET calculations (bottom panel), respectively. These are cumulative
distributions for all momentum and atmospheric depth bins. Each ratio is 
weighted by the square of the error on the 
ratio derived from the experimental error on the intensity. The distribution
for 1D calculations has a mean 1.13, indicating an average overprediction
of 13\%, whereas for 3D calculations the mean is 1.07, a slightly better
agreement. The main improvement however is in the reduced RMS variance of 
0.17 for the distribution for 3D calculations compared to 0.27 for the 1D 
calculations. Thus, the more realistic calculations that take into account 3D 
air shower development and primary zenith arrival direction constitute a clear
improvement in the representation of muon production.

\subsubsection{Neutrino Production}

\begin{figure}
\epsfig{file=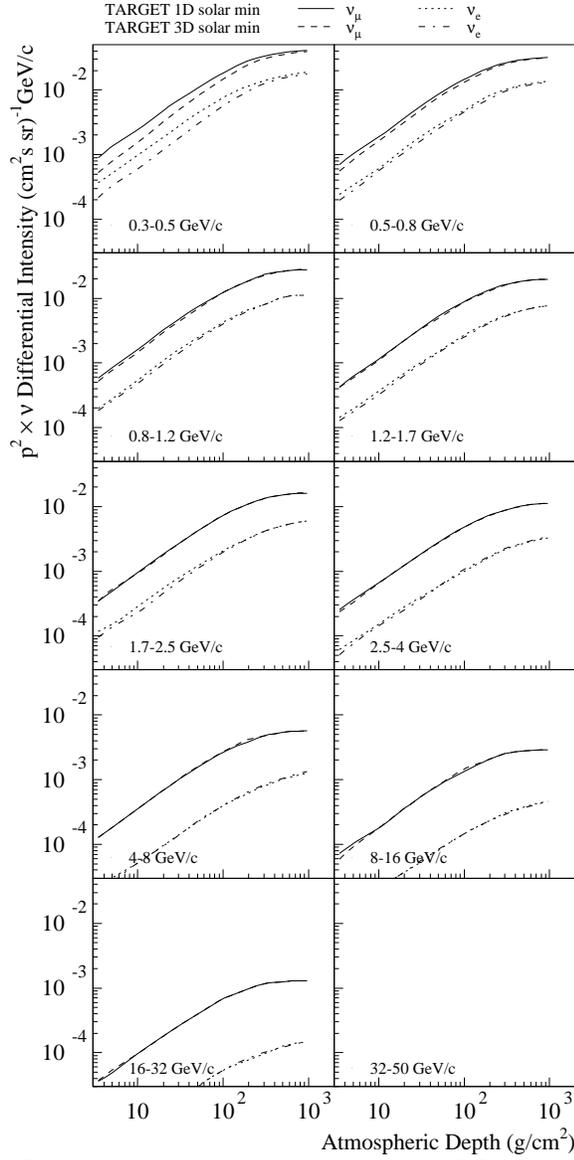,width=8cm}
\caption{
$\nu$ growth curves predicted by TARGET at Lynn Lake, for solar minimum
conditions, in different momentum intervals. The curves are:
1D $(\nu_\mu + {\overline \nu}_\mu)$, solid curves;
3D $(\nu_\mu + {\overline \nu}_\mu)$, dashed curves;
1D $(\nu_{\rm e} + {\overline \nu}_{\rm e})$, dotted curves;
3D $(\nu_{\rm e} + {\overline \nu}_{\rm e})$, dot-dashed curves.
\label{nuflx}}
\end{figure}

The 1D and 3D TARGET algorithms were used to predict ($\nu_\mu + 
{\overline \nu}_\mu$) and ($\nu_{\rm e} + {\overline \nu}_{\rm e}$)
intensities at different atmospheric depths and in different momentum
bins. The calculations are made for solar minimum conditions, for
no geomagnetic rigidity cutoff, and for primary cosmic rays arriving
within 30$^\circ$ of the zenith. The resulting neutrino growth curves
in different momentum intervals are shown in Figure~\ref{nuflx}. The
calculations are made only up to 32 GeV/c. Although there are differences 
between the 1D and 3D predictions at momenta less than about 1~GeV/c, these 
differences are most important at mid-to-high altitudes. The neutrino 
intensities at the ground level, which are the ones of relevance to the 
underground neutrino studies, are summarized in Table~\ref{nuflxgnd}, for
1D and 3D calculations. The neutrino intensities at ground level appear not 
to be altered much by the 3D effects. A similar conclusion was also reached
in a study by Battistoni {\it et al.} \cite{bat:nu}, where detailed 
calculations
of atmospheric muon and neutrino production are made in one and three 
dimensions.

\section{Conclusions}

We have made statistically significant measurements of air shower muons as 
a function of atmospheric depth. 
We report the muon charge ratio $\mu^+/\mu^-$ in the momentum range 
0.3-0.9~GeV/c and the momentum spectra of $\mu^-$ in the range 0.3-50~GeV/c, 
at atmospheric depths from 13 to 960~g/cm$^2$.
The charge ratio is essentially constant with altitude within
errors, with a possible decrease from 1.3-1.4 at high altitudes to 
1.1 at the ground. A comparison of our measured $\mu^-$ momentum distributions 
with model calculations indicates significant discrepancies 
with the predictions of the standard one-dimensional TARGET algorithm:
our measured fluxes are lower than the calculated ones at shallow depths
before about shower maximum. Calculations of the muon intensities with a new 
version of the TARGET algorithm, accounting for three-dimensional air shower 
development, lead to a substantially improved agreement with our data. A 
detailed representation of atmospheric secondary production thus benefits from 
the more realistic simulations. The average excess of about 7\% of the 3D 
calculations over our measured intensities is comparable to possible 
systematic effects in our experiment. Thus, within this uncertainty, the 3D 
TARGET algorithm generates atmospheric secondary particle intensities which
are in agreement with the measurements.

The three-dimensional air shower development effects do
not appear to impact significantly the atmospheric neutrino rates at the 
ground, but merely the pattern of neutrino production altitudes. Thus,
we estimate that the neutrino intensities predicted by the 1D version
of the TARGET algorithm are also accurate to about 7\%. This is to be
compared with the accuracy of 14-18\% first estimated for such calculations
\cite{agr:nu}. Even though the Honda {\it et al.} \cite{hon:nu} model uses
different assumptions about the primary cosmic ray flux and about the
atmospheric interaction characteristics, it predicts neutrino intensities
on the ground which agree with the TARGET predictions and with our data
at the level of 7-10\%. Therefore, one might put into question
the additional normalization factor of 1.16 of the primary cosmic ray spectrum 
that is introduced in the SuperKamiokande neutrino oscillation analysis.

\acknowledgments

We are grateful to T. Stanev and T. K. Gaisser for helpful discussions, 
for sharing with us their TARGET Monte Carlo algorithm and for contributing
to the effort to introduce 3D effects in the algorithm.
We thank the NSBF balloon crews that have supported the HEAT flights.
This work was supported by NASA grants NAG5-5059, NAG5-5069, NAG5-5070, 
NAGW-5058, NAGW-1995, NAGW-2000 and NAGW-4737, and by financial assistance 
from our universities.

\appendix

\vskip 4cm

\newpage

.

\newpage

\begin{table}
\caption{Muon counts at $0.3\leq\left|R\right|\leq0.9$~GV as a function of
atmospheric depth $d$. Uncertainties are statistical. \label{mupmum}}
\begin{tabular}{ccccc}
 $d$ &  $\overline{d}$ &  N$_{\mu^+}$ &  N$_{\mu^-}$ &  $\mu^+/\mu^-$ \\
 (g/cm$^2$)  &  (g/cm$^2$) &              &              &  \\
\tableline
  3-4   &  3.45 &   134 &    65 &  $2.06\pm0.31$ \\
  4-7   &  5.13 &    35 &    27 &  $1.30\pm0.33$ \\
  7-13  &  9.85 &    53 &    37 &  $1.43\pm0.31$ \\
 13-32  &  23.2 &   305 &   193 &  $1.58\pm0.15$ \\
 32-67  &  49.1 &   458 &   348 &  $1.316\pm0.094$ \\
 67-140 &  105  &  1264 &   954 &  $1.325\pm0.057$ \\
 140-250 &  190  &  1691 &  1463 &  $1.156\pm0.041$ \\
 250-350 &  298  &   605 &   540 &  $1.120\pm0.066$ \\
 350-840 &  499  &   810 &   792 &  $1.023\pm0.051$ \\
 840-960 &  957  &  1076 &   990 &  $1.087\pm0.048$ \\
\end{tabular}
\end{table}

\begin{table}
\caption{Rigidity-dependent parameters and corrections used in the 
determination of the absolute $\mu^-$ intensities.\label{reffic}}
\begin{tabular}{ccccc}
$R$ & $\overline{R}$ & $\epsilon_{dt}$ & $(\Omega A)$ & $\epsilon_{acc}$ \\
 (GV) &  (GV) &  (\%) &  (cm$^2$sr) &  (\%) \\
\tableline
  0.3-0.5 &  0.40 &  68.35$\pm$0.22 &  568.9$\pm$2.6 &  52.0$\pm$0.5 \\
  0.5-0.8 &  0.65 &  74.46$\pm$0.20 &  613.1$\pm$2.8 &  63.0$\pm$0.5 \\
  0.8-1.2 &  0.99 &  78.77$\pm$0.19 &  608.2$\pm$2.7 &  66.1$\pm$0.5 \\
  1.2-1.7 &  1.43 &  80.46$\pm$0.18 &  604.4$\pm$2.7 &  65.4$\pm$0.5 \\
  1.7-2.5 &  2.06 &  81.57$\pm$0.18 &  600.0$\pm$2.7 &  64.6$\pm$0.5 \\
  2.5-4   &  3.13 &  81.64$\pm$0.18 &  601.9$\pm$2.7 &  63.9$\pm$0.5 \\
  4-8     &  5.52 &  81.48$\pm$0.18 &  598.9$\pm$2.7 &  62.5$\pm$0.5 \\
  8-16    &  11.0 &  81.45$\pm$0.18 &  603.6$\pm$2.7 &  61.2$\pm$0.5 \\
  16-32   &  21.3 &  81.45$\pm$0.18 &  606.0$\pm$2.7 &  47.0$\pm$0.4 \\
  32-50   &  39.2 &  81.43$\pm$0.18 &  602.4$\pm$2.7 &  15.5$\pm$0.2 \\
\end{tabular}
\end{table}

\begin{table}
\caption{Depth-dependent parameters and corrections used in the determination 
of the absolute $\mu^-$ intensities.\label{deffic}}
\begin{tabular}{cccccc}  
 $d$ &  $\overline{d}$ &  $\Delta t$ &  $\epsilon_t$ &  $\epsilon_l$ &  
$\epsilon_\theta$ \\
 (g/cm$^2$) &  (g/cm$^2$) &  ($\pm 1$ s) &  (\%) &  (\%) &  (\%) \\
\tableline 
 13-32 &  23.2 &  1453 &  48.1$\pm$1.8 &  89.190$\pm$0.079 &   99.68 \\
 32-67 &  49.1 &  1307 &  41.9$\pm$1.6 &  96.585$\pm$0.047 &   99.33 \\
 67-140 &  105 &  1571 &  49.5$\pm$4.6 &  95.051$\pm$0.051 &   98.58 \\
 140-250 &  190 &  1285 &  65.8$\pm$6.9 &  92.894$\pm$0.072 &   97.45 \\
 250-350 &  298 &  484 &  82.9$\pm$0.3 &  95.45$\pm$0.12 &   96.05 \\
 350-840 &  499 &  1163 &  95.7$\pm$0.2 &  94.30$\pm$0.14 &   93.50 \\
 840-960 &  957 &  5335 &  99.3$\pm$0.1 &  97.09$\pm$0.10 &   88.03 \\
\end{tabular}
\end{table}

\newpage

.

\newpage

\widetext
\begin{table}
\caption{Number of recorded $\mu^-$ events and their intensity (in
(cm$^2$s sr GeV/c)$^{-1}$) as a function of momentum and atmospheric depth.
\label{fluxes}}
\begin{tabular}{cccccc}  
{} &  0.40 GeV/c &  0.65 GeV/c &  0.99 GeV/c &  1.43 GeV/c &  2.06 GeV/c \\
\hline 
 23.2 &  81 &  86 &  89 &  42 &  51 \\
 g/cm$^2$ &  $(4.12\pm0.67)\times 10^{-3}$ &  $(2.06\pm0.33)\times 10^{-3}$  & 
           $(1.45\pm0.23)\times 10^{-3}$   &  $(5.5\pm1.1)\times 10^{-4}$ & 
           $(4.22\pm0.77)\times 10^{-4}$ \\
 49.1 &  135 &  181 &  100 &  97 &  90 \\
 g/cm$^2$ &  $(6.7\pm1.0)\times 10^{-3}$ &  $(4.42\pm0.62)\times 10^{-3}$   & 
           $(1.68\pm0.26)\times 10^{-3}$ &  $(1.29\pm0.20)\times 10^{-3}$ & 
           $(7.6\pm1.2)\times 10^{-4}$ \\
 105 &  354 &  494 &  403 &  285 &  241 \\
 g/cm$^2$ &  $(1.32\pm0.20)\times 10^{-2}$ &  $(8.7\pm1.3)\times 10^{-3}$   & 
           $(4.86\pm0.75)\times 10^{-3}$ &  $(2.74\pm0.43)\times 10^{-3}$ & 
           $(1.46\pm0.23)\times 10^{-3}$ \\
 190 &  568 &  721 &  626 &  441 &  370 \\
 g/cm$^2$ &  $(2.02\pm0.32)\times 10^{-2}$ &  $(1.21\pm0.19)\times 10^{-2}$  & 
           $(7.2\pm1.1)\times 10^{-3}$ &  $(4.03\pm0.65)\times 10^{-3}$ & 
           $(2.13\pm0.34)\times 10^{-3}$ \\
 298 &  198 &  268 &  245 &  191 &  189 \\
 g/cm$^2$ &  $(1.47\pm0.20)\times 10^{-2}$ &  $(9.4\pm1.3)\times 10^{-3}$   & 
           $(5.84\pm0.79)\times 10^{-3}$ &  $(3.63\pm0.51)\times 10^{-3}$ & 
           $(2.26\pm0.31)\times 10^{-3}$ \\
 499 &  235 &  441 &  398 &  309 &  275 \\
 g/cm$^2$ &  $(6.68\pm0.89)\times 10^{-3}$ &  $(5.83\pm0.73)\times 10^{-3}$  & 
           $(3.56\pm0.45)\times 10^{-3}$ &  $(2.21\pm0.29)\times 10^{-3}$ & 
           $(1.23\pm0.16)\times 10^{-3}$ \\
 957 &  280 &  550 &  624 &  630 &  653 \\
 g/cm$^2$ &  $(1.85\pm0.23)\times 10^{-3}$ &  $(1.61\pm0.19)\times 10^{-3}$  & 
           $(1.22\pm0.14)\times 10^{-3}$ &  $(9.8\pm1.2)\times 10^{-4}$ & 
           $(6.36\pm0.75)\times 10^{-4}$ \\
\end{tabular}
\begin{tabular}{cccccc}  
{} &  3.13 GeV/c &  5.52 GeV/c &  11.0 GeV/c &  21.3 GeV/c &  39.2 GeV/c \\
\hline 
 23.2 &  34 &  23 &  7 &  4 &  0 \\
 g/cm$^2$ &  $(1.52\pm0.32)\times 10^{-4}$   & 
           $(4.04\pm0.97)\times 10^{-5}$ &  $(6.5\pm2.6)\times 10^{-6}$   & 
           $(2.4\pm1.2)\times 10^{-6}$  &  {}                            \\
 49.1 &  62 &  44 &  22 &  4 &  1 \\
 g/cm$^2$ &  $(2.80\pm0.49)\times 10^{-4}$   & 
           $(8.0\pm1.5)\times 10^{-5}$ &  $(2.02\pm0.49)\times 10^{-5}$   & 
           $(2.6\pm1.3)\times 10^{-6}$ &  $(1.8\pm1.8)\times 10^{-6}$  \\
 105 &  173 &  128 &  44 &  12 &  3 \\
 g/cm$^2$ &  $(5.73\pm0.94)\times 10^{-4}$   & 
           $(1.68\pm0.29)\times 10^{-4}$ &  $(2.93\pm0.61)\times 10^{-5}$   & 
           $(5.3\pm1.7)\times 10^{-6}$ &  $(3.6\pm2.1)\times 10^{-6}$  \\
 190 &  276 &  216 &  77 &  14 &  2 \\
 g/cm$^2$ &  $(8.7\pm1.4)\times 10^{-4}$   & 
           $(2.69\pm0.45)\times 10^{-4}$ &  $(4.86\pm0.93)\times 10^{-5}$   & 
           $(6.3\pm1.9)\times 10^{-6}$ &  $(2.7\pm1.9)\times 10^{-6}$  \\
 298 &  151 &  126 &  40 &  10 &  2 \\
 g/cm$^2$ &  $(9.9\pm1.4)\times 10^{-4}$   & 
           $(3.25\pm0.48)\times 10^{-4}$ &  $(5.2\pm1.0)\times 10^{-5}$   & 
           $(8.8\pm3.0)\times 10^{-6}$ &  $(4.9\pm3.5)\times 10^{-6}$  \\
 499 &  272 &  247 &  95 &  30 &  4 \\
 g/cm$^2$ &  $(6.61\pm0.87)\times 10^{-4}$   & 
           $(2.35\pm0.31)\times 10^{-4}$ &  $(4.57\pm0.71)\times 10^{-5}$   & 
           $(9.9\pm2.1)\times 10^{-6}$ &  $(3.5\pm1.8)\times 10^{-6}$  \\
 957 &  723 &  735 &  380 &  124 &  13\\
 g/cm$^2$ &  $(3.79\pm0.45)\times 10^{-4}$   & 
           $(1.49\pm0.17)\times 10^{-4}$ &  $(4.11\pm0.50)\times 10^{-5}$   & 
           $(8.6\pm1.2)\times 10^{-6}$ &  $(2.34\pm0.70)\times 10^{-6}$  \\
\end{tabular}
\end{table}

\begin{table}
\caption{Near vertical neutrino intensities at 960~g/cm$^2$ at zero 
geomagnetic cutoff, for 1D and 3D TARGET calculations. All intensities
are in (cm$^2$ s sr GeV/c)$^{-1}$.\label{nuflxgnd}}
\begin{tabular}{ccccc}
 Momentum & $(\nu_\mu + {\overline \nu}_\mu)$  & $(\nu_\mu + 
{\overline \nu}_\mu)$ & 
$(\nu_{\rm e} + {\overline \nu}_{\rm e})$ & $(\nu_{\rm e} + 
{\overline \nu}_{\rm e})$ \\
 (GeV/c)  & 1D Intensity & 3D Intensity & 1D Intensity & 3D Intensity \\
\tableline
 0.36  & 0.32 & 0.31 & 0.15 & 0.14 \\
 0.71  & 0.063 & 0.062 & 0.027 & 0.026 \\
 0.89  & 0.035 & 0.035 & 0.014 & 0.014 \\
 1.4   & 0.0099 & 0.0099 & 0.0038 & 0.0038 \\
 1.8   & 0.0051 & 0.0052 & 0.0019 & 0.0019 \\
 2.8   & 0.0014 & 0.0014 & $4.1 \times 10^{-4}$ & $4.3 \times 10^{-4}$ \\
 5.6   & $1.8 \times 10^{-4}$ & $1.8 \times 10^{-4}$ & $4.0 \times 10^{-5}$ 
& $4.2 \times 10^{-5}$ \\
 11    & $2.3 \times 10^{-5}$ & $2.3 \times 10^{-5}$ & $3.7 \times 10^{-6}$ 
& $3.6 \times 10^{-6}$ \\
 22    & $2.6 \times 10^{-6}$ & $2.6 \times 10^{-6}$ & $2.9 \times 10^{-7}$ 
& $3.0 \times 10^{-7}$ \\
\end{tabular}
\end{table}


\begin{references}

\bibitem{fuk:osc} Y. Fukuda {\it et al.}, {\sl Phys. Rev. Lett.} {\bf 81},
1562 (1998).

\bibitem{bar:nu} G. Barr, T. K. Gaisser, and T. Stanev, {\sl Phys. Rev.} D 
{\bf 39}, 3532 (1989).

\bibitem{lee:nu} H. Lee and Y. S. Koh, {\sl Nuov. Cim.} B {\bf 105}, 
883 (1990).

\bibitem{hon:nu} M. Honda {\it et al.}, {\sl Phys. Lett.} B {\bf 248}, 
193 (1990), M. Honda {\it et al.}, {\sl Phys. Rev.} D {\bf 52}, 4985
(1995).

\bibitem{kaw:nu} M. Kawasaki and S. Mizuta, {\sl Phys. Rev.} D {\bf 43}, 
2900 (1991).

\bibitem{agr:nu} V. Agrawal {\it et al.}, {\sl Phys. Rev.} D {\bf 53}, 
1314 (1996).

\bibitem{bes:pfl} T. Sanuki {\it et al.}, {\sl Proceedings of the 
26th International Cosmic Ray Conference, Salt Lake City, 1999} 
(University of Utah, Salt Lake City, 1999), 
Vol. 3, p. 93.

\bibitem{gai:nu} T. K. Gaisser, T. Stanev, and G. Barr, {\sl Phys. Rev.} D 
{\bf 38}, 85 (1988). G. Barr, T. K. Gaisser, and T. Stanev, {\sl Phys. Rev.}
D {\bf 39}, 3532 (1989).

\bibitem{gai:rat} T. K. Gaisser {\it et al.}, {\sl Phys. Rev.} D {\bf 54},
5578 (1996).

\bibitem{mas:mu} R. Bellotti {\it et al.}, {\sl Phys. Rev.} D {\bf 53}, 
35 (1996).

\bibitem{mas2:mu} R. Bellotti {\it et al.}, {\sl Phys. Rev.} D {\bf 60}, 
052002 (1999).

\bibitem{sch:mu} E. Schneider {\it et al.}, {\sl Proceedings of the 
24th International Cosmic Ray Conference, Rome, 1995} (ARTI Grafiche 
Editoriali SRL, Ubino, 1995), Vol. 1, p. 690.

\bibitem{heat:mu} G. Tarl\'e {\it et al.}, {\sl Proceedings of the 
25th International Cosmic Ray Conference, Durban, 1997} (Potchefstroomse
Universiteit, Potchefstroomse, South Africa, 1997), Vol. 6, p. 321.
S. Coutu {\it et al.}, {\sl Proceedings of the 29th International
Conference on High Energy Physics, Vancouver, 1998} (World Scientific,
Singapore, 1999), Vol. 1, p. 666.

\bibitem{boe:mu} M. Boezio {\it et al.}, {\sl Phys. Rev. Lett.} {\bf 82},
4757 (1999).

\bibitem{kri:mu} J. F. Krizmanic {\it et al.}, {\sl Proceedings of the 
24th International Cosmic Ray Conference, Rome, 1995} (Ref. \cite{sch:mu}),
Vol. 1, p. 593. J. F. Krizmanic (private communication).

\bibitem{bar:nim} S. W. Barwick {\it et al.}, {\sl Nucl. Inst. \& Meth.} 
A {\bf 400}, 34 (1997).

\bibitem{fas:flu} A. Fass\` o {\it et al.}, {\sl Proceedings of the
IVth International Conference on Calorimetry in High Energy Physics, 
La Biodola, 1993} (World Scientific, Singapore, 1994).

\bibitem{ste:sec} S. A. Stephens, {\sl Proceedings of the 17th International 
Cosmic Ray Conference, Paris, 1981} (Commissariat \'a l'Energie Atomique,
Paris, 1981), Vol. 4, p. 282.

\bibitem{bas:mu} G. Basini {\it et al.}, {\sl Proceedings of the 24th 
International Cosmic Ray Conference, Rome, 1995} (Ref. \cite{sch:mu}),
Vol. 1, p. 585.

\bibitem{zen:cor} Y. L. Blokh {\it et al.}, {\sl Nuov. Cim.} B
{\bf 37}, 198 (1977).

\bibitem{bat:nu} G. Battistoni {\it et al.}, {\it Astropart.
Phys.} {\bf 12}, 315 (2000).

\end{references}
\end{document}